\documentclass[nofootinbib,twocolumn]{revtex4-1} 
\usepackage{amsmath}  % needed for \tfrac, \bmatrix, etc.
\usepackage{amsfonts} % needed for bold Greek, Fraktur, and blackboard bold
\usepackage{graphicx} % needed for figures
\usepackage{hyperref}

\begin{document}

\title{Variability as a better characterization of Shannon entropy}

\author{Gabriele Carcassi}
\email{carcassi@umich.edu}
\author{Christine A. Aidala}
\email{caidala@umich.edu}
\affiliation{Department of Physics, University of Michigan, Ann Arbor, MI 48109-1040 }
% Please provide a full mailing address here.

\author{Julian Barbour}
\email{julian.barbour@physics.ox.ac.uk}
\altaffiliation[permanent address: ]{College Farm, South Newington, Banbury, Oxon, OX15 4JG UK}
\affiliation{Visiting Professor in Physics at the University of Oxford (2009-2018)}

\date{\today}

\begin{abstract}
	The Shannon entropy, one of the cornerstones of information theory, is widely used in physics, particularly in statistical mechanics. Yet its characterization and connection to physics remain vague, leaving ample room for misconceptions and misunderstanding. We will show that the Shannon entropy can be fully understood as measuring the variability of the elements within a given distribution: it characterizes how much variation can be found within a collection of objects. We will see that it is the only indicator that is continuous and linear, that it quantifies the number of yes/no questions (i.e.~bits) that are needed to identify an element within the distribution, and we will see how applying this concept to statistical mechanics in different ways leads to the Boltzmann, Gibbs and von Neumann entropies.
\end{abstract}

\maketitle

%\tableofcontents

\section{Introduction\label{int}}

The goal of this work is to give a comprehensive introduction to the concept of Shannon entropy, the expression $-\sum p_i \log p_i$, which is well established in such disparate fields as computer science, communication theory, physics \cite{Jaynes1957-1, Grandy} ecology and economics. The concept of entropy in physics, despite being treated extensively in the physics education literature (e.g. \cite{baierlein1994entropy, carson2002undergraduate, Santillan_2008, LeffPartI, LeffPartII, LeffPartIII, LeffPartIV, LeffPartV, Benguigui_2013, geller2014entropy, Tiwary_2020}), is still surrounded by a great deal of confusion \cite{Swendsen, Styer}. Partly because, at a fundamental level, there is still disagreement on what is the ``correct'' notion of entropy, particularly out of equilibrium \cite{jarzynski2017stochastic, goldstein2019gibbs, maes2003time}. Partly because the same word ``entropy'' is used to refer to concepts (e.g. topological entropy, thermodynamic entropy, graph entropy, R\'{e}nyi entropy, information entropy) that have different definitions and various degrees of overlap.  We will focus on the notion of Shannon entropy, by which we mean the use of $- \sum_i p_i \log p_i$ in the broadest characterization, which has received less attention in physics education. We will see that the Shannon entropy can be given a crisp and precise characterization and will explore its relation to the Gibbs and Boltzmann entropies in statistical mechanics.\footnote{We will not discuss its relation to thermodynamic entropy. Thermodynamic entropy is defined on the notions of heat and work while the Shannon entropy is defined on the notion of a distribution. The contexts are different and relating them goes beyond the scope of this work. } The same material is also presented in a series of videos \cite{videos}.

To give a crisp and intuitive meaning to the Shannon entropy we need to address two main problems. The first problem is a consequence of its success. As it is used in many fields, it is generally introduced with concepts that are specific to that field but may not be appropriate for others. As we will see, the meaning of the $p_i$ can vary significantly. The first goal, then, is to provide a characterization of the Shannon entropy that satisfies the following requirements:
\begin{itemize}
	\item it is defined in a general setting that is crisp and intuitive
	\item it is precise enough so that the formula itself can be derived
	\item it recovers the specific meaning when applied to each field.
\end{itemize}
In this paper we show that the idea of variability of the elements within a distribution, how diverse the objects are from each other within the collection, can serve as such a general concept. We will therefore use the term \emph{``Shannon variability''} for this more general idea that applies to all cases, leaving ``entropy'' to the more specific physical concept. We will show that the expression $- \sum_i p_i \log p_i$ is the only linear indicator of variability and it measures the number of questions one must ask to identify an element of the distribution, linking its use to information theory. The variability of messages within a communication channel will correspond to its information entropy as used in communication theory; the variability of microstates within a macrostate will correspond to the Gibbs entropy used in statistical mechanics; the variability of the state of a single particle within a microstate will correspond to the Boltzmann entropy. This avoids the use of incomplete or imprecise characterizations (e.g. uncertainty, knowledge, lack of knowledge, disorder) that, while useful in some limited circumstances, promote confusion in others.

The second problem is that the Shannon formula presents some peculiarity when applied over continuous variables that is often the source of confusion. Namely, it is generally linked to a choice of unit, of the resolution used to describe the problem. However, the physical state spaces, in both classical and quantum mechanics, provide additional structure that make the Shannon variability coordinate-independent. This is seldom pointed out, and it is critical to form a proper connection to the Boltzmann, Gibbs and von Neumann entropies.

While some details presented here may be well known in a particular community, we find that they may be new to another. Therefore we assume little knowledge of the subject from the reader and include all details that we feel are important to properly understand the subject without confusion, including two standard derivations for the formula and examples that are common in computer science but rarely used in physics. We will briefly touch on a few of the common misconceptions and show how they are resolved.

\section{Variability within a Distribution\label{vwd}}

The general setting is the following. We have a set of elements $E = \{e_\alpha\}_{\alpha=1}^N$. In accordance with the comments made in the introduction, they can for example represent the galaxies in the Laniakea Supercluster, the animals in the Gal\'{a}pagos islands, the molecules in a box of gas given a particular macrostate, the different outcomes of spin measurement for a given quantum state, or the words in the complete works of Shakespeare. The choice of the particular set is driven by the interests and practices of scientists and their fields of study and therefore it is, in this sense, discretionary. Yet, it must be objective in at least one respect: once a choice is made the actual elements are the same for everybody. Once we arbitrarily decided we will study the galaxies in the Laniakea Supercluster, it is a matter of fact that the Milky Way will be included. This also means that, whatever properties those elements have, they will represent a matter of fact about that set. In turn, this will make some choices arguably more appropriate or insightful than others. For example, in taxonomy the set of all animals that have feathers may be more appropriate than the set of all animals that can fly: the first includes only birds while the second includes most birds, a few mammals and a lot of insects.

Once a set of elements is chosen, we select a property or a set of properties we want to use to characterize the elements. That is, we have a set of possible values $Q=\{q_i\}_{i=1}^I$ and a map $q : E \to Q$ that associates a value to each element. These can represent the galaxy types, the genus of the animal, the possible states of the molecules, the possible values for spin or the spelling of words. This will give us a sequence $\{q(e_\alpha)\}_{\alpha=1}^N$ of the descriptions that are associated with each element in the set. Again, the choice of properties is ultimately discretionary, grounded on what particular aspects of the elements we are studying in a given case. Yet, like before, once a choice is made the value for each element is objectively defined. Once we arbitrarily decided we are studying galaxy types it is a matter of fact that the Milky Way is a barred spiral galaxy. Again, one may find that some choices are arguably more appropriate or interesting than others.

Having decided what elements to study and the level of description in which we are interested, we bin them; that is, we group them based on that description, disregarding the identity of the particular element. What we will be interested in is only the relative frequency $p_i = N_i / N = \left| \{e_\alpha \in E | q(e_\alpha) = q_i \} \right| / N$ of the elements within each bin. We may do so because we are either not interested or not able to further distinguish the elements. Whatever the reason, once the choices of elements and properties are made, this relative frequency is objective. The nature of the $p_i$ will depend on the previous choices. It may represent the fraction of elements if the set is constituted by a group of objects. It may represent a probability of an outcome if the set consists of different realizations of similarly prepared systems. We will therefore call $p_i$ weights, to be clear that we make no commitment as to whether we have fractions, frequentist probabilities, Bayesian probabilities or other notions.

Once we have constructed our distribution, we want to construct an indicator $H$ that quantifies how much variability the elements exhibit within the distribution. That is, we want to quantify the degree of diversity that the values can have within the distribution. To that end, we want to define some suitable requirements for $H$.

We may be tempted to use standard statistical quantities, like the range or the variance, but this is not possible. First, if the values associated with the bins are non-numerical (e.g. types of galaxies, words), such statistical quantities are not well defined. Second, relabeling the values (e.g.~switching names, switching units, non-linearly changing coordinates) does not have an impact on variability of the elements, while statistical quantities will in general be affected. This tells us that our indicator cannot depend on $Q$ itself, but only on the weights $p_i$. That is, we require $H=H(p_i)$.\footnote{Note that for continuous quantities the weights are densities and are affected by the choice of $Q$: the unit is required to specify the numeric value (e.g.~$1 \% / $mm) and this will change under unit transformation (e.g.~$1000 \% /$m). Therefore, as the weights $p_i$ themselves depend on the unit, the indicator $H(p_i)$ will in general depend on the choice of $Q$. Though this is a source of additional confusion, for the purpose of defining a measure of variability it does not change things conceptually. We discuss the issue in section \ref{sec_cont}.}

We  expect small changes in a distribution -- small changes of the weights -- to produce small changes in variability; we therefore require $H$ to be a continuous function of the $p_i$. We also expect that as the number of values found within the distribution increases, so will the variability. Therefore if we have a uniform distribution over $I$ cases, so that $p_i = 1/I$, we require $H$ to be monotonically increasing with $I$.

As noted before, the level at which we describe the elements is not absolute and can change. For example, we may choose to group the animals in the Gal\'{a}pagos islands first by class (e.g.~mammals, birds, reptiles) and then later refine the mammals by species. In this case, we would like $H$ to combine linearly with respect to the weights. That is, we want the variability of the overall distribution $H_T = H_C + p_M H_M$ to be the variability over the classes $H_C$ plus the variability of the mammals $H_M$ weighted by the fraction of mammals $p_M$. This also makes the quantity additive when combining two independent distributions.

To sum up, we have the following three requirements:
\begin{enumerate}
	\item $H$ depends only on $p_i$ and it does so continuously
	\item If $p_i=1/I$ then $H$ is a monotonically increasing function of $I$
	\item Let $p_i$ and $q_j$ be the weights for two distributions respectively over $I$ and $J$ bins. Let $r_k$ be the distribution over $K=I+J-1$ bins constructed by expanding the $a^{th}$ bin of the first distribution using the second distribution. More specifically, let $1 \leq a \leq I$, then $r_k = \{p_1, p_2, ..., p_{a-1}, p_{a}q_1, p_{a}q_2, ..., p_{a}q_J, p_{a+1}, ..., p_I \}$. Then $H(r_k) = H(p_i) + p_{a} H(q_j)$.
\end{enumerate}
These are the same requirements Shannon put forth for his expression\cite{Shannon}, from which he showed that the only possible choice is $H(p_i) = - \sum p_i \log(p_i)$. See also \cite{khinchin2013mathematical} for a similar derivation. This expression quantifies the variability of the elements within the distribution, the variety of values one finds. The precise meaning of this variability is context dependent, as the choices of the elements and binning are not fixed and the meaning of the weights depends on what the distribution is describing. But this is true for any mathematical object: a real number may represent mass, color in the frequency spectrum, the total money supply, the half-life for an isotope, a probability and so on.

The Shannon variability may represent uncertainty in some cases, if the weights are probabilities or credences, but not in the general case. If the weights portray the fraction of the elements that have a certain property, like the fraction of galaxies in the Laniakea Supercluster that are barred spirals, there is nothing uncertain about the distribution. In this case, the Shannon variability represents how much variation we have within galaxies in terms of galaxy types.

The Shannon variability may represent knowledge in some cases, but not in general. Consider the following two cases:
\begin{enumerate}
	\item There is 50\% chance you won one million dollars in the lottery and 50\% chance you won nothing.
	\item There is 50\% chance you won one million dollars in the lottery and 50\% chance you won half a million dollars.
\end{enumerate}
The distribution in both cases is the same, two bins 50\% chance each, and so is the Shannon variability. Yet, you know more in the second case: you know you won at least half a million dollars.

Unfortunately, entropy in general is often associated with vague characterizations like the two presented. It is said to represent uncertainty, knowledge, lack of knowledge or disorder depending on the authors, which leads to confusion and misunderstanding.  The characterization we have given of the Shannon variability measure, on the other hand, applies to all cases and leads naturally to the assumptions required to rederive it. Our characterization therefore is more fundamental. If we look at the galaxies in our universe, what variability do they exhibit in terms of their types? If we look at the animals in the Gal\'{a}pagos islands, what variability is expressed in terms of their species? If we look at the molecules in a given macrostate, what variability do they express in terms of their microstates? If we look at the words in the complete works of Shakespeare, what variability do we find in his vocabulary?

If the Shannon expression is a measure of variability, why is it connected to information? How is variability quantified and in what units?

\section{Units of Variability\label{uv}}

To understand what the numerical value represents, consider this example. Suppose we fix a distribution, say the animals in the Galap\'{a}gos islands binned by their respective species. Suppose we pick a specific animal from the set and you want to know its species. Suppose the only way for you to get that information is to ask a series of questions with only two possible answers, yes or no. How many questions would you have to ask? In other words, we are playing a game of Twenty Questions.

Not all questions will be able to extract the same amount of information. Some questions, like, ``Is it an animal?" would be redundant. Others, like, ``Is it an American Flamingo (Phoenicopterus ruber)?", would give us a lot of information in the positive case but little in the negative case. However, there has to be a minimum number of questions that must be asked to get to the answer. A single question, for example, cannot be enough given that there are more than two species. It should be intuitively clear that to a greater variability within the distribution will correspond a greater number of questions you must ask. That is exactly what the Shannon variability quantifies: the minimum average number of questions one has to ask to identify a value in the distribution. It gives us the number of questions for an ideal strategy for our game of Twenty Questions.

If we have binary questions, the logarithms will be in base two and the unit for Shannon variability will be bits. It will indicate the average number of yes/no questions we need to identify an element within the distribution. In general, you can pick any base $b$: in base three we have ternary questions and trits, for ten we have questions with ten possible answers and digits. We can also pick a non-integer base, like the natural base for logarithms, and we will have nats. This is why the Shannon variability is fundamental in information theory, because it quantifies \emph{how much information is needed to transfer a value picked from a known distribution}.

Defining a set of questions means choosing an encoding as we are choosing how the information gets codified into our series of bits. To understand how this works, we can briefly review the Huffman coding,\cite{Huffman} which is the optimal algorithm for symbol-by-symbol coding with a known probability distribution. The idea is that we want all possible answers to each question to be balanced, to provide the same amount of information. The reason is that making one answer more specific (i.e.~it applies in fewer cases) means making another less specific (i.e.~it applies in more cases). In the case of binary questions, then, we ideally want the probability to answer yes or no to be 50\%.

For example, suppose the population of pets in a country is as follows:
\begin{table}[h]
	\begin{tabular}{lr}
		dogs & 27\% \\
		cats & 48\% \\
		fish & 10\% \\
		birds & 8\% \\
		small mammals & 4\% \\
		reptiles & 3\%
	\end{tabular}
\end{table}

\noindent
For the first question, we group cats on one side and everything else on the other, to form a 48/52 split. So we can ask, ``Is it a cat?". If the answer is yes, we finished. If not, we need to continue. We can group dogs on one side and everything else on the other to form a 27/25 split. So we can ask, ``Is it a dog?". If the answer is yes, we are done. If not, we continue. We can group fish with reptiles and birds with small mammals to form a 13/12 split. So we can ask, ``Is it a fish or a reptile?". If the answer is yes, the followup question would be ``Is it a fish?". If the answer is no, the followup question would be ``Is it a bird?". With this scheme, the encoding, where 1 represents 'yes' and 0 represents 'no' to the each of the questions asked, is as follows:
\begin{itemize}
	\item dogs 27\% - 2 questions (i.e.~2 bits) - answers: [no, yes] (i.e.~encoding 01)
	\item cats 48\% - 1 question (i.e.~1 bit) - answers: [yes] (i.e.~encoding 1)
	\item fish 10\% - 4 questions (i.e.~4 bits) - answers: [no, no, yes, yes] (i.e.~encoding 0011)
	\item birds 8\% - 4 questions (i.e.~4 bits) - answers: [no, no, no, yes] (i.e.~encoding 0001)
	\item small mammals 4\% - 4 questions (i.e.~4 bits) - answers: [no, no, no, no] (i.e.~encoding 0000)
	\item reptiles 3\% - 4 questions (i.e.~4 bits) - answers: [no, no, yes, no] (i.e.~encoding 0010)
\end{itemize}
The number of questions needed in each case corresponds to the number of bits. The answers in each case are represented by the encoding. Note how the encoding depends on the specific choice of questions. We can calculate the average number of bits for the encoding to be:  $(.48) * 1 + (.27) * 2 + (.1 + .08 + .04 + .03) * 4 = 2.02$ bits. This represents the average number of bits we would have to use for each animal if we repeated the game many times. We can also calculate the Shannon variability to be $.27 * \log_2(.27) + .48 * \log_2 (.48) + ... =1.98$ bits. This represents the ideal case, the minimum number of questions required to reach a definite answer. Note that our encoding is already very close to the ideal case.

Now that we understand that the Shannon variability is measured by the number of bits required to identify an element from a distribution, it is instructive to derive the same expression from different considerations. Suppose we have a sequence of $N$ elements, say pets like in the previous example. These are taken from $I$ different cases: dogs, cats, fish and so on. Suppose $N_i$ are the number of elements of each type, which means $\sum_{i \in I} N_i = N$. Then, given a particular instance of $N_i$, we have $W(N_i) = \frac{N!}{\prod_{i \in I} N_i!}$ possible ways to realize that sequence, which corresponds to all possible permutations. If all permutations are equally likely, then $\log W (N_i)$ represents the number of bits needed to identify one of the sequences.

When $N$ is large, we can use Stirling's approximation $\ln N! \approx N \ln N - N$, and find $\log W(N_i) \approx - N \sum_{i \in I} \frac{N_i}{N} \log \frac{N_i}{N} = N H(p_i)$. The logarithm of the permutations is $N$ times the Shannon variability. The result should not be surprising: it simply tells us that encoding a sequence of $N$ elements is the same as encoding $N$ elements one at a time.

It is important, at this point, to understand that the technical use of the term information in information theory does not equate to the normal use of the term which refers to knowledge, intelligible data. The bits by themselves, the yeses and the nos, the ones and the zeros, do not provide knowledge. They need the context of the questions and the distribution to become actual information. For example, when opening a jpeg file, the file itself does not contain the instructions of how to read it. If you do not happen to know what a jpeg is and how to read it, you are not going to be able to interpret, to decode, the string of bits into actual intelligible data. The questions, the distribution, the context are considered given, communicated out-of-band through another scheme. As with any semantic content, this cannot be easily formalized and quantified.

The Shannon variability of the distribution, then, has nothing to do with the information the distribution itself holds. The distribution is not what is being encoded. The variability is quantified by the information needed to go from the distribution, which is given, to an individual element. It is really the information gap from the population to an element. That is why some people say the entropy is ``lack of information," which is justified because, in a way, it is the information about the elements that the distribution cannot provide. But, again, this is deceptive: if one is not interested in identifying elements there is no ``missing information."

In communication and information theory, information is really encoded information. Communication systems and information processors have no idea whether the source of the data is a digital thermometer or a poet. There is no knowledge per se, just symbol manipulation that may represent different concepts in different contexts. The term information entropy, then, is misleading for two reasons. First, because it is really not entropy in the thermodynamic sense: it is not defined on states, it does not know about irreversible processes and it is unrelated to maximization at equilibrium. In fact, it is not related to physical systems. Second, it is not really information in the general sense, only in the very narrow technical sense of encoded information within a communication or information system.

The use of information in physics, then, does not warrant a fundamental change of perspective in what constitutes a physical object, as some physicists have claimed. It is true that any physical process can be used to process information, \emph{when properly encoded}. It is also true that scientific theories, in the end, are models that can capture only the aspects of nature that can be tested experimentally, the information extracted by the experiment, under suitable circumstances. Therefore the claim that information plays an essential role in physical theories has a valid basis.\footnote{Comments like ``It is wrong, moreover, to regard this or that  physical quantity as sitting \emph{out there} with this or that numerical value'', ``the information thus solicited [by the experiment] makes physics and comes in bits'' by Wheeler\cite{Wheeler} or ``I am proposing that the ultimate form of the implementable laws of physics requires only operations available (in principle) in our actual universe'' by Landauer\cite{Landauer} go in this direction} But it is also true that that data requires the context in order to be understood: we need to know what the subject of our experiment is, how to prepare it and how to collect the data. The art of experimental science is contained neither in the mathematical description nor in the data collected. As information, in the information theoretic sense, requires that context to become intelligible, it cannot play a primary role. Therefore the claim that the universe itself is information\footnote{Comments like ``[Information is] the fundamental building block of the Universe'' by Vedral\cite{Vedral} or ``[Information] occupies the ontological basement'' by Davies\cite{Davies} go in this direction.} does not follow.

We have seen what the Shannon variability measures and why it is important in information theory, but we have so far worked with discrete quantities. In physics we are also interested in continuous quantities, like position and momentum. How does it work in that case? Would we not need infinitely many bits to identify a value from a continuous distribution?

\section{Continuous Distributions}\label{sec_cont}

A standard quick and dirty way to extend distributions from discrete variables to continuous variables is to substitute weights with densities and sums with integrals. Therefore, instead of having a discrete normalized distribution $\sum_i p_i = 1$, we have a continuous normalized distribution $\int \rho(x)dx=1$. While simply changing $- \sum_i p_i \log p_i$ to $- \int \rho(x) \log \rho(x) dx$ works in most cases, leaving it at that misses a crucial point: the two expressions have significantly different properties.

In the discrete case, the Shannon variability is always positive. In the continuous case, the Shannon variability can be negative. For example, consider a uniform distribution over a line:
\begin{equation}
\rho(x)=
\begin{cases}
0 & x < a\\
\dfrac{1}{b-a} & a \leq x \leq b\\
0 & b < x
\end{cases}
\end{equation}
We have $H(\rho) = - \int_a^b \frac{1}{b-a} \log \frac{1}{b-a} dx = \log (b-a)$. If $0 < b-a < 1$ the Shannon variability will be negative. What does it mean to have a negative number of bits?

When considering continuous distributions, we have to remember that the densities are given per unit. That is, $\rho(x)$ has dimensions $[\frac{1}{x}]$. The variability will be also given relative to the unit. For example, a uniform distribution over a unit interval will correspond to zero variability. A uniform distribution over a two unit interval will correspond to one bit, since we will need one bit to narrow the variability back to one unit. A distribution over half a unit interval will correspond to minus one bit, since we would need to ``lose'' one bit of information to widen the variability back to one unit. In other words, the variability is measured by the number of bits needed to identify an element up to one unit, up to the resolution (i.e. the granularity) given by that unit.

This means that infinitesimally narrow distributions, like delta functions, do not work well with the Shannon variability as they would give minus infinite entropy. If one wants entropy over continuous variables to be bounded, then one must work with continuous functions which  maybe have very small but still finite support (i.e.~region where the function is non-zero), which is what we will assume.

This brings up a second problem: what happens if we change units? If we change the reference, we expect that the entropy will change accordingly. In fact we find
\begin{equation}
\begin{aligned}
H(\rho(y)) &= H(\rho(x)) - \int \rho(x)  \log \left|\frac{\partial y}{\partial x}\right|  dx 
\end{aligned}
\end{equation}
In general, the Shannon variability over a continuous variable is not invariant under coordinate transformations. Under translations, the Jacobian $\left|\frac{\partial y}{\partial x}\right|$ is unitary, so the Shannon variability does not change. If we stretch or shrink, if we change scale, the Shannon variability changes to measure the variability at the new scale. If the change is non-linear, the variability over different ranges will be counted differently.

This may seem particularly perplexing given that we want to use this quantity in a physical setting. If we want $\rho(x)$ to represent the state of a macroscopic system and $H(\rho(x))$ to represent entropy, a state variable, how can $H$ change value if we merely express $\rho$ over different coordinates? How can maximization of entropy be meaningful if it yields different results depending on the coordinate system? Does that mean that we can use the Shannon formula only on distributions over discrete values?

In physics, what makes the Shannon variability work is phase space. What happens is that, under coordinate transformation $\hat{q}^i = \hat{q}^i(q^j)$, the $d\hat{q}^i = \frac{\partial \hat{q}^i}{\partial q^j } dq^j$ vary like vector components while $d\hat{p}_i = \frac{\partial q^j}{\partial \hat{q}^i } dp_j$ vary like covector components. Therefore the areas $d\hat{q}^i d\hat{p}_i = dq^j dp_j$ remain the same. This has a couple of important consequences. First of all, (1) the volumes are preserved during the transformation, which means (2) the Jacobian is unitary and (3) the Shannon variability is invariant. In fact, these three properties are mathematically the same property: each one implies all others. Second, something similar happens to the areas for each independent degree of freedom, meaning that the Shannon variability associated to the marginal distributions is also invariant. These properties can be taken to be the defining characteristics of phase space: phase space has exactly such geometrical properties.\cite{AoP2020HamiltonianEntropy}

The use of phase space in statistical mechanics is of paramount importance: over this space both the density $\rho(q^i, p
_i)$ and its variability $H(\rho(q^i, p
_i))$ are invariant under arbitrary transformations of $q^i$, including non-linear ones. This is not just a coincidence or convenience: it is essential if we want to give the density at each point of phase space a physically objective character.

The importance of phase space is often attributed to Liouville's theorem, which states that areas in phase space are conserved under Hamiltonian evolution. While this is true, their invariance under coordinate transformations is more important as it is what makes them physically meaningful in the first place. This means that we always have to consider distributions over the full phase space with the appropriate conjugate coordinates if we want to use $-\int \rho \log \rho$. If we use distributions over position and velocities or over momentum only, the expression will change for those particular variables in those coordinate systems. This can be, again, a source of confusion.\cite{Dunkel}

We conclude noting that in quantum mechanics the Shannon variability, in the form of the von Neumann entropy, is coordinate invariant as well. That is, given a density matrix $\rho$ then $S_\textrm{\tiny{vN}} = - \textrm{tr}(\rho \log \rho)$ is independent of the basis in which it is calculated.

Now that we have seen how the Shannon variability works in the continuous case and the importance of phase space, we are ready to see how it can be used in statistical mechanics and how it relates to the Boltzmann, Gibbs and von Neumann entropies.

\section{Connection to Statistical Mechanics\label{csm}}

Statistical mechanics aims to describe the collective behavior of a large number of physical systems (i.e.~particles), therefore asking what variability is expressed by such a collection is a well posed question. Yet, we have to understand that there are two distinct ways to apply the concept simply because there are two types of distributions in which one can be interested. We may consider the state of the whole system at a given time, and be interested in the distribution of the particles over all possible particle states. This will lead to the Boltzmann entropy. We may instead consider a statistical ensemble, which is a large collection of independent copies of the system found in different states, and consider the distribution of these different instances over the states of the whole system. This will lead to the Gibbs or the von Neumann entropy depending on whether the system is classical or quantum. Both these types of distributions are used and are therefore of interest, so we will examine both.

Suppose we have a large number $N$ of particles taken from a normalized distribution $\rho(q^i, p_i)$. The space is the six-dimensional phase space for a single particle, sometimes called $\mu$-space. As we assume $N$ large, we can think of $N\int_\Sigma \rho(q^i, p_i)dq^idp_i$ as the number of particles that are within a region $\Sigma$ of phase space. That is, $\rho$ does not represent a probability distribution but an actual physical distribution that tells us the state of the whole system at one instant of time. The Boltzmann entropy is given by $S_B = k_B \log W$, where $W$ will correspond to the different ways that the $N$ particles can be arranged while still satisfying the distribution. If the space were discrete, the computation of the permutations would be straightforward. But how does this work in a continuous space?

As we said before, we treat continuous variables by comparing to a finite unit. We can pick a unit of phase space small enough such that the density $\rho$ can be considered constant over cells of that size. We express $\rho$ as the number of particles within the chosen unit and divide phase space into cells. Now we can calculate all the possible permutations of the particles within the different cells and, in these circumstances, we will find that $\log W = N H(\rho)$. That is, the number of permutations at that level of precision will be equivalent to the number of particles times the variability of the distribution at that level of precision. In other words, the Boltzmann entropy reduces to $N$ times the Shannon variability of the single-particle distribution for a given microstate.

% http://przyrbwn.icm.edu.pl/APP/PDF/120/a120z6p04.pdf

The Boltzmann constant $k_B$ should not distract us: its role is simply to allow us to measure temperature in an appropriate unit.\footnote{In the latest revision of the International Metric System of Units SI,\cite{chyla2011evolution} the Boltzmann constant is one of the defining constants. This means its value is assigned, rather than measured, in a way that properly defines Kelvin, the unit of temperature. The universal constants are essentially used as fixed starting points to define our measurement scales.
}
If one defines $T = \frac{\partial U}{\partial S}$, and measures temperature in Kelvin and energy in Joules, the above relationship forces us to measure entropy in Joules/Kelvin. Therefore the entropy cannot be expressed as a pure number. However, this is not the only possible definition. Instead of using $T$ as a primary thermodynamic variable, we can use $\beta = \frac{1}{k_B T}$. In this case, one would define $\beta = \frac{\partial S}{\partial U}$, measure energy in Joules, $\beta$ in inverse Joules and entropy would be dimensionless. As one can do all the thermodynamic calculations using just $\beta$, the constant $k_B$ is really just set by the unit system.\footnote{As \cite{chyla2011evolution} notes:
	\begin{quote}
		Temperature characterizes the average thermal energy of particles in a certain ensemble of particles in the state of thermodynamic equilibrium. Therefore, in principle, one could express this quantity in terms of the unit of energy; this is actually practiced in many fields of physics, especially in statistical physics, where it is quite common to use $\Theta = k_B T$ or $\beta = 1 / k_B T$  as a measure of thermal energy, instead of temperature $T$ measured in Kelvins.
	\end{quote}
}

Distributions over single-particle phase space constitute the setting used when deriving the Maxwell-Boltzmann distribution $\rho(q^i, p_i) = \left(\frac{\beta m}{2\pi} \right)^{3/2}e^{-\beta \frac{p_ip^i}{2m}}$, which is the distribution of particles for an ideal gas. As it evolves towards equilibrium, particles will spread out as much as they can under the constraints given by the energy, volume and number of particles, increasing the variability until it is maximized. The equilibrium is a statistical equilibrium, particles are moving around, but for any particle that moves in one direction, there is another one that moves in the opposite and the overall distribution remains the same.\footnote{This corresponds to the original insights developed by Boltzmann.}

This approach, though, will only work in the limit of a fixed large number of indistinguishable particles that are independently distributed. More precisely, suppose we have a joint probability distribution $\hat{\rho}$ for $N$ particles. If they are independently distributed, then the joint probability $\hat{\rho}=\prod_{i}\rho_i$ is the product of the distribution for each particle. If they are indistinguishable, then $\rho_i=\rho$: each particle has the same distribution. If $N$ is large, the number of particles in one region $N_\Sigma$ is very close to the expectation value $N\int_\Sigma \rho(q^i, p_i)dq^idp_i$. If these assumptions are not met, we cannot break the joint distribution into single-particle ones, particle number and type may change, the fluctuations may become relevant thus requiring a more general account.\footnote{Jaynes\cite{Jaynes} has shown that a single distribution over $\mu$-space will not recover the correct experimental values for entropy.}

The more general setting, then, is the following. The macroscopic state, or macrostate, is a probability distribution over all possible complete descriptions of the system, or microstates. That is, we have a distribution over the $6N$-dimensional phase space of $N$ particles, sometimes referred to as $\Gamma$-space, where each point represents the position and momentum of $N$ particles. The Gibbs entropy is $S_G = -k_B \int \rho \log \rho \, dq^idp_i$, which corresponds to the Shannon variability of the microstate distribution for a given classical macrostate. The Gibbs entropy, then, is the variability of a microstate as it moves around within the macrostate. The macrostate of an equilibrium will be fully identified by a set of macroscopic variables, such as temperature, average energy, pressure, and so on. Note that these may be quantities that are not defined on an individual microstate but only on the ensemble. The microscopic dynamics will be free to move around as long as those statistical quantities are preserved. The Gibbs entropy, then, tells us the variability of the microstate under the given constraints and, at equilibrium, we will find that variability to be maximal.

There are a couple of issues in this picture. The first is that $\Gamma$-space automatically assumes that all particles are distinguishable. This leads to the widely known problem of overcounting which needs to be addressed in the standard way. The second problem is that, though the state of each particle is given by position and momentum, we should not think of them as literally pointlike. As we said, this would correspond to delta Dirac distributions over phase space which have minus infinite Shannon variability. It is more appropriate, both mathematically and conceptually, to think of particles as identically peaked distributions, each characterized by the same amount of Shannon variability.\footnote{Setting to $-\log h$ the entropy corresponding to each degree of freedom of these peaked distributions is a natural way to incorporate the effects of the uncertainty principle.} The position and momentum correspond to the center of mass of the particle.

We note that some authors choose to interpret the probability distribution not as coming from repeated independent trials, but as the knowledge one has about the system. This would make the entropy a subjective notion: each observer would have a different credence distribution, regardless of whether it fits the data, and therefore a different Shannon entropy. A physical quantity, however, must be the same for everybody, and the associated fluctuations we experimentally observe are indeed objective. So what is going on?

We believe the confusion comes from the, correct, realization that the same system, under different conditions, will be described with a different set of thermodynamic variables. \footnote{Jaynes\cite{Jaynes} points out:
	\begin{quote}
		Consider, for example, a crystal of Rochelle salt. For one set of experiments on it, we work with temperature, pressure, and volume. The entropy can be expressed as some function $S_e(T,P)$. For another set of experiments on the same crystal, we work with temperature, the component $e_{xy}$ of the strain tensor, and the component $P_z$ of electric polarization; the entropy as found in these experiments is a function $S_e(T,e_{xy},P_z)$. It is clearly  meaningless to ask, ``What is the entropy of the crystal?''  unless we first specify the set of parameters which define its thermodynamic state.
\end{quote}}
The choice of the system, in statistical mechanics and thermodynamics, is enough to determine the state space for the microstates, but not enough to determine the set of macrostates that correspond to equilibria. We have to specify the process and the constraints that that process puts on the system. Under a different choice of process and constraints the microstates will fluctuate in different ways since we have changed the dynamics of the system. This is what we stated at the beginning: a distribution, and therefore its variability, is always contingent upon some arbitrary choices. Since the Gibbs entropy is the variability of a microstate within the distribution identified by the macrostate, it is not a property of the single microstate, it is not a property of the system, but it is a property of the system within that specific process, of the macrostate. Some authors refer to this fact by saying that entropy is not objective \footnote{Jaynes\cite{Jaynes,Jaynes2} called it the anthropomorphic nature of entropy.}. 

The issue is that subjective, in the context of probability theory, refers to Bayesian probability, which really means subjective: what one believes to be true regardless of what experimental evidence there is. This is not at all the same concept as the one outlined before.\footnote{Note how Jaynes\cite{Jaynes2} always put ``subjective'' in double quotes.} The system plus the process (which identifies the set of constraints) determines the ensemble and therefore the entropy, regardless of whether an agent knows what process was used. Even if one wants to give a Bayesian account, then, one has to give it in terms of an agent that has full and exact knowledge of the system and the process. But this is hardly subjective. In our view, this is a case of unfortunate word choice. It is much better to simply state that the entropy is a property of the ensemble, of the distribution, and not of the system. And if one understands that the Gibbs entropy is the variability of the microstate within the ensemble, then it is clear that it cannot be a property of the microstate itself, but it is a property of the ensemble.

The case of a quantum system is formally similar to the classical one. Instead of a distribution over the $N$-particle phase space, we have a distribution over the Hilbert space for the quantum system which is represented by a density matrix operator $\rho$. The von Neumann entropy is given by $S_\textrm{\tiny{vN}} = - \textrm{tr}(\rho \log \rho)$. This, expanded in a basis, becomes $S_\textrm{\tiny{vN}} = - \sum_i \rho_i \log \rho_i$ or $S_\textrm{\tiny{vN}} = - \int \rho(x) \log \rho(x)dx$ depending on whether the spectrum is discrete or continuous.  The von Neumann entropy corresponds to the Shannon variability of the microstate distribution for a given quantum macrostate.

The different entropies in statistical mechanics, then, all have a tight link to the Shannon variability. The Boltzmann entropy corresponds to the variability of the state of a particle within a given microstate, provided that there are a large fixed number of independently distributed and indistinguishable particles. The Gibbs entropy corresponds to the variability of a classical microstate as constrained by the macrostate. The von Neumann entropy is similar but corresponds to the quantum case. The characterization we gave to the Shannon formula, then, is readily applicable to statistical mechanics in a natural way. 

\section{Conclusion}

In this paper we have seen that:
\begin{itemize}
	\item the Shannon entropy measures the variability of the elements within a given distribution, giving it a crisp intuitive meaning that is general and applicable to all branches of science
	\item the expression is not arbitrary, as it is the only linear indicator for such a concept
	\item it measures the variability by quantifying the number of yes/no questions one must ask to identify an element within the distribution, which corresponds to the number of bits needed to transmit or store that information
	\item when properly applied to statistical mechanics, the variability leads to the Boltzmann, Gibbs and von Neumann entropies.
\end{itemize}
The characterization we gave to the Shannon formula, then, is more precise than the common characterizations, such as disorder, information or lack of knowledge, and it should lead to less confusion. It clarifies that the Shannon variability is an independent concept from the entropy of thermodynamics and statistical mechanics, and a link can be recovered only if properly applied. We find that this approach, once internalized, gives greater intuitive insight and also maps more readily to the mathematical details.

\section{Acknowledgments}
We would like to thank Juniar Lucien for insights into the physics education literature. G.C. and
C.A.A. acknowledge funding from the MCubed program
of the University of Michigan. This work is in connection to Assumptions of Physics, a larger project that aims to identify a handful of physical principles from which the basic laws can be rigorously derived  (\url{https://assumptionsofphysics.org}).

\bibliographystyle{unsrt}  
\bibliography{bibliography}

\end{document}